\documentclass[lettersize,journal]{IEEEtran}
\usepackage{amsmath,amsfonts}
\usepackage{algorithmic}
\usepackage{algorithm}
\usepackage{array}
\usepackage[caption=false,font=normalsize,labelfont=sf,textfont=sf]{subfig}
\usepackage{textcomp}
\usepackage{stfloats}
\usepackage{url}
\usepackage{verbatim}
\usepackage{graphicx}
\usepackage{cite}
\usepackage{upgreek}
\usepackage{xcolor}
\usepackage{amsmath}
\usepackage{bm}
\renewcommand{\algorithmicrequire}{\textbf{Input:}}
\renewcommand{\algorithmicensure}{\textbf{Output:}}

\begin{document}

\title{Decoupling Spatio-Temporal Dynamics: Microvibration Imaging Using Coherent Detection Ghost Imaging Lidar}

\author{Shuang Liu, Jinquan Qi, Chaoran Wang, Chenjin Deng, Shensheng Han
\thanks{Shuang Liu, Jinquan Qi, Chaoran Wang, Chenjin Deng, Shensheng Han are with the Shanghai Institute of Optics and Fine Mechanics, Chinese Academy of Sciences (e-mail:liushuang0820@siom.ac.cn).}
}



\maketitle

\begin{abstract}
Imaging the full-field microvibration of extended targets remains a formidable challenge for conventional remote sensing. Traditional array-based sensors are often severely constrained by data throughput and sensitivity limits when scaling to high spatial resolutions, while point-scanning interferometric systems lack the instantaneous full-field capability required to capture transient, spatially coupled vibration modes. To overcome these limitations, we propose a Coherent Detection-Ghost Imaging (CD-GI) framework that synergizes the spatial multiplexing capability of single-pixel imaging with the high-dimensional sensitivity of coherent detection. We establish a comprehensive mathematical model that describes the coupling mechanism of the target’s spatial distribution and temporal micro-dynamics within a 1D bucket detector signal. To resolve the resulting inverse problem, we develop a frequency-channel self-calibration scheme. This approach effectively decouples the micro-Doppler signatures from spatial speckle patterns without requiring prior knowledge of the vibration frequency. Experimental results demonstrate that our system successfully reconstructs the spatially resolved microvibration patterns of adjacent targets with a frequency difference as small as 1 Hz, achieving sub-wavelength vibration sensitivity. This work bridges the gap between computational imaging and coherent metrology, offering a robust solution for non-invasive, high-precision structural health monitoring.

\end{abstract}


\section{Introduction}
\IEEEPARstart{M}{ICRO} vibration full-field imaging, often referred to as structural acoustic field imaging, has emerged as a critical technology for non-invasive diagnostics in high-stakes sectors, including aerospace structural health monitoring, precision manufacturing optimization, and biomedical tissue characterization\cite{muellerComparingGRACEFOKBR2022}. Unlike static imaging, micro-vibration imaging requires the simultaneous acquisition of high-resolution spatial structures and transient temporal dynamics. This capability is essential for detecting early-stage mechanical fatigue, characterizing resonant modes in micro-electromechanical systems (MEMS), and identifying pathological tissue properties. However, capturing these high-dimensional spatio-temporal dynamics with high fidelity remains a significant challenge, as it demands detection systems that are simultaneously sensitive to sub-wavelength displacements and capable of instantaneous full-field acquisition.

Current approaches typically suffer from a fundamental trade-off between spatial resolution, motion sensitivity, and system complexity. Microwave radar systems, particularly Inverse Synthetic Aperture Radar (ISAR), have developed sophisticated algorithms for micro-Doppler analysis\cite{zhangRemovalMicroDopplerEffect2021,zhangMicroDopplerEffectsRemoved2021}. However, they are fundamentally limited by their long wavelengths, lacking the micro-Doppler sensitivity required to resolve micrometer-scale vibrations\cite{chenMicrodopplerEffectRadar2006,junDetectionMicroDopplerEffect2014,lachettaSimulatingDigitalMicromirror2021}. Conversely, optical interferometric techniques, such as Laser Doppler Vibrometry (LDV) \cite{rothbergInternationalReviewLaser2017} and Polytec Scanning Vibrometers (PSV) \cite{vasiliuBasePlateResonance2024}, offer exquisite sub-nanometer displacement sensitivity. However, these systems fundamentally operate as zero-dimensional point detectors. To image an extended target, they must rely on sequential raster scanning, which introduces inevitable temporal latencies. This scanning mechanism not only precludes the capture of transient or non-repeatable phenomena but is also susceptible to mechanical jitter and synchronization errors\cite{halkonEstablishingCorrectionSolutions2021,venkatakrishnanTwoaxisscanningLaserDoppler2002,zhuNumericalJitterEstimation2020}. While Focal Plane Array (FPA) sensors can achieve snapshot imaging, scaling coherent detection arrays to high resolutions imposes prohibitive costs and generates unmanageable data throughput, creating a bottleneck for real-time sensing.

To overcome these limitations, we propose a CD-GI framework that synergizes the spatial multiplexing advantages of computational ghost imaging \cite{shapiro2012physics,qiu2023remote,zeng2025tailoring,erkmen2012computational} and video snapshot compressive imaging\cite{yuanPlugandPlayAlgorithmsLargeScale2020} with the high detection sensitivity of coherent lidar \cite{ip2008coherent,taylor2009phase,kikuchi2015fundamentals,fink1975coherent}. Unlike scanning-based methods, CD-GI encodes the full-field target information into a single-pixel bucket detector using structured light illumination. By employing a coherent detection scheme, the system retains the phase information essential for micro-Doppler extraction. However, a critical challenge arises: the bucket detector integrates signals from all spatial points, causing the temporal vibration signatures to couple inextricably with the spatial speckle patterns. To resolve this ill-posed problem without prior knowledge of the vibration frequency, we introduce a frequency-channel self-calibration scheme. This approach allows us to virtually "synchronize" the random vibration phases from different pulses, enabling the coherent integration and decoupling of specific vibration modes from the composite signal.

The main contributions of this paper are summarized as follows:
\begin{itemize}
\item We establish a comprehensive spatio-temporal coupling model for coherent ghost imaging, mathematically describing how the target’s spatial distribution and temporal micro-dynamics are encoded into a one-dimensional coherent signal.
\item We develop a frequency-channel self-calibration algorithm that effectively decouples vibration information from spatial information. This algorithm enables the reconstruction of spatially resolved vibration modes even for non-cooperative targets with unknown frequencies.
\item We experimentally demonstrate the system’s capability to resolve and image distinct microvibration patterns of adjacent targets with a frequency separation of only 1 Hz, achieving robust performance in sub-wavelength vibration sensing.
\end{itemize}

The remainder of this paper is organized as follows. Section II details the system architecture of the CD-GI lidar and derives the mathematical model characterizing the spatio-temporal coupling mechanism. It also presents the formulation of the frequency-channel self-calibration algorithm. Section III provides a theoretical analysis of the frequency selectivity and validates the proposed method through numerical simulations. Section IV describes the experimental setup and reports the results of the full-field microvibration mode reconstruction. Section V discusses the current limitations and potential improvements, and Section VI concludes the paper.

\begin{figure*}[t!]
\centering
\includegraphics[width=1.86\columnwidth,height=0.5\linewidth]{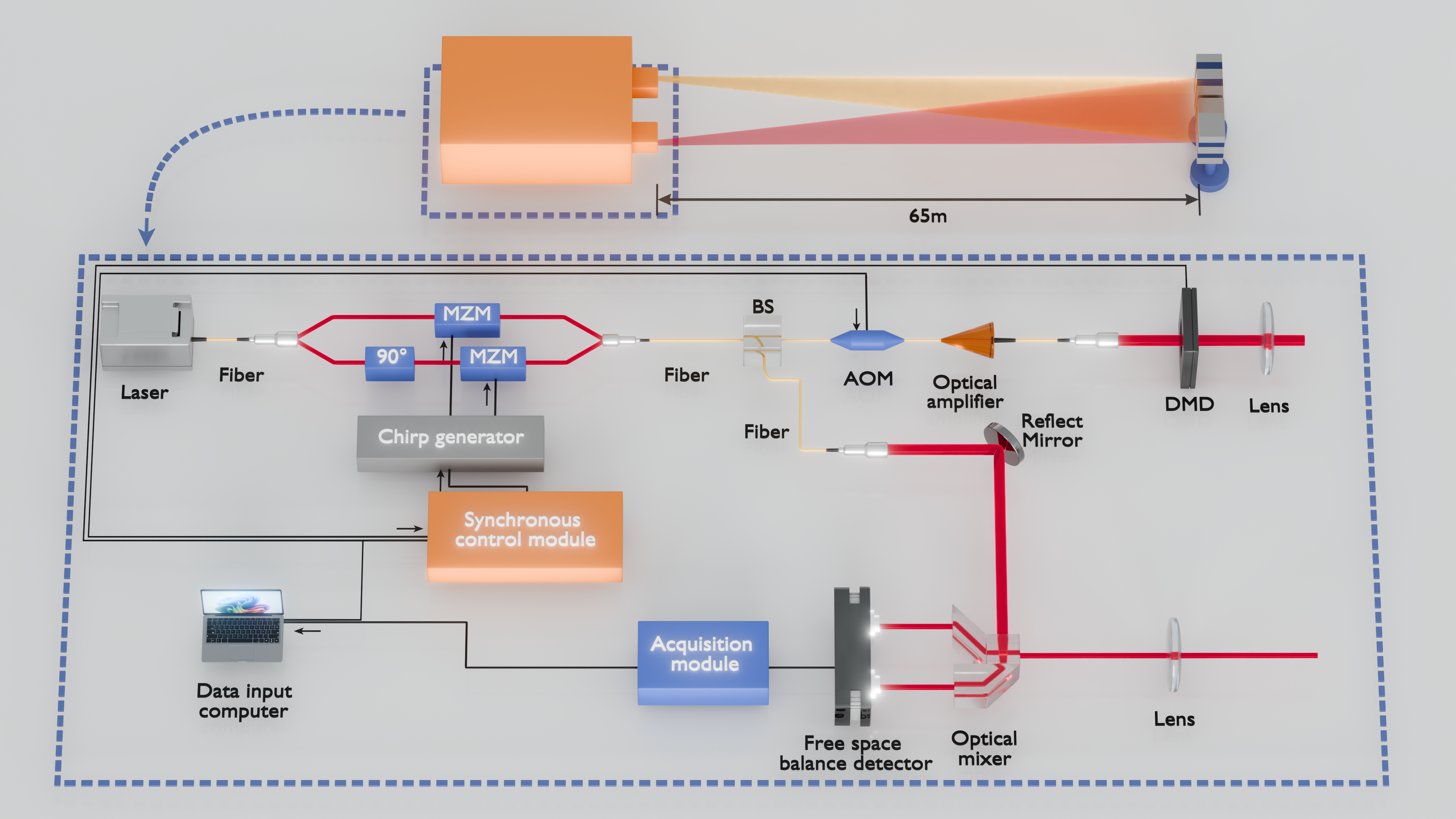}
\caption{Schematic diagram of the proposed CD-GI lidar system. The architecture is based on a Mach-Zehnder interferometer where the transmitting path is frequency-shifted by an AOM and spatially modulated by a DMD. The backscattered echo interferes with the $LO$ at a balanced detector, converting the high-dimensional spatio-temporal target information into a one-dimensional heterodyne signal.}
\label{system_structure}
\end{figure*}

\section{SYSTEM ARCHITECTURE AND SIGNAL MODELING}

\subsection{SYSTEM ARCHITECTURE}
CD-GI Lidar Architecture The schematic of the proposed CD-GI lidar system is illustrated in Fig. \ref{system_structure}. The architecture is based on a Mach-Zehnder interferometer configuration employing a narrow-linewidth laser source at 1550 nm. To achieve high-resolution ranging and velocity discrimination, the source is phase-modulated to generate a Frequency-Modulated Continuous Wave (FMCW) signal. 
The modulated beam is split into a reference path (Local Oscillator, $LO$) and a transmitting path. In the transmitting path, the optical carrier is frequency-shifted by an intermediate frequency (IF), denoted as $f_{IF}$, via an acousto-optic modulator. Crucially, to enable single-pixel imaging, a Digital Micromirror Device (DMD) is placed in the transmitting path to impose time-varying spatial modulation patterns, $P^i(\rho_o)$, onto the beam, where $i$ denotes the pulse index and $\rho_o$ represents the spatial coordinates on the target plane.\\

\begin{figure*}[t!]
\centering
\includegraphics[width=1.7\columnwidth,height=0.43\linewidth]{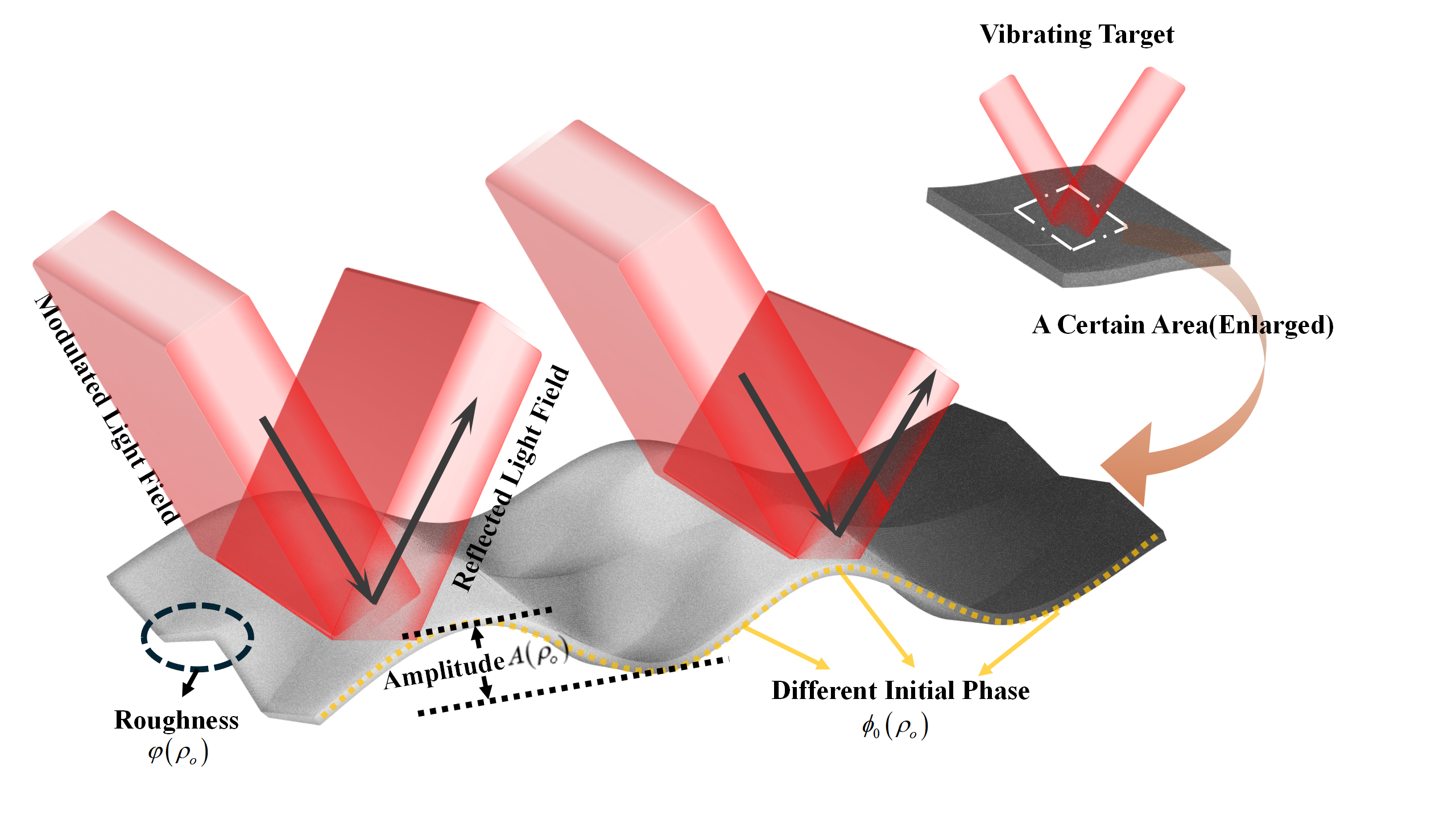}
\caption{A magnified view of a specific probing area on the target surface during detection. The initial phase distribution and amplitude profile across the target surface collectively constitute the vibration mode morphology of the target.}
\label{target_surface}
\end{figure*}

The backscattered echoes from the target are collected by the bucket detector and mixed with the $LO$ beam on a spatial optical mixer. A balanced photodetector then converts the optical interference field into a one-dimensional electrical signal. This signal preserves both the spatial information encoded by the DMD and the temporal Doppler signatures induced by the target's microvibrations.

\subsection{Signal Model}
In our previous work\cite{liuMicrovibrationModesReconstruction2022}, we demonstrated a reconstruction method for a cooperative vibrating target, where the pulse repetition frequency (PRF) of the detecting light field can be set as an integer multiple of the vibration frequency. Due to the spatiotemporal characteristics of the target's microvibration mode response, it is essential to ensure that the initial phase of the probe pulses is synchronized with the starting moment of the target's vibration cycle. This alignment enables coherent integration of the echoes over multiple vibration periods, thereby achieving sufficient signal-to-noise ratio (SNR) for subsequent signal processing. The interaction with a vibrating surface and the physical process of signal acquisition are described in this work\cite{qiMultimodeCoherentDetection2026}. The full complex detected optical field at the receiver plane can be expressed as:
\begin{equation}
\label{qi_work}
\begin{aligned}
E^{i}_d(t, \rho_d)& \propto \varphi_d(\rho_d) \cdot s(t - \tau_0) \cdot \left[ \Psi^{i} * h_{sys} \right] (\rho_d)\\
& \times \exp[j2\pi (f_c+f_{IF})t] \cdot \exp(j\phi_i).
\end{aligned}
\end{equation}
The physical interpretations of each term in Eq. \ref{qi_work} are described as follows: 
\begin{itemize}

    \item [1)] $\rho_d$ denotes the coordinates of the receiver plane. $i$ is the serial number of the pulse. $t$ denotes the intra-pulse sampling sequence, also referred to as the fast-time. Thus, $E^{i}_d(t, \varphi_d)$ represents the temporal evolution of the optical field on the receiver surface during the $i$-th pulse.\\
    \item[2)] $\varphi_d(\rho_d)$ represents the parabolic quadratic phase term in the Fresnel diffraction formulation, evaluated at the detector plane.\\
    \item[3)] $s(t - \tau_0)$ is the received signal, which can be designed according to detecting requirements. And $\tau_0$ is the time delay.\\
    \item[4)] $\Psi^i$ represents the spatially modulated light field projected onto the target, which carries the encoding pattern $P^i(\rho_o)$ generated by the DMD. $h_{sys}$ denotes the impulse response of the free-space propagation channel. Consequently, $\left[ \Psi^{i} * h_{sys} \right] (\rho_d)$ represents the convolution between the encoded target and receiving module's point spread function (PSF). Noticing that the intensity patterns are selected to modulate the complex-valued target in this work. \\
    \item[5)] $\exp[j2\pi (f_c+f_{IF})t]$ represents the modulation with the carrier wave whose frequency is $f_c$ and the IF signal whose frequency is $f_{IF}$.\\
    \item[6)] $\exp(j\phi_{i})$ accounts for global phase perturbations, such as atmospheric turbulence, or the platform-induced motion.
    
\end{itemize}

In the reference path, the beam serves as the $LO$, which can be modeled as a plane wave $E_{LO} \propto \exp(j2\pi f_{c}t)$. The reflected echo $E^{i}_d(t, \rho_d)$ from the transmitting path and the $LO$ beam interfere on the surface of the balanced detector. The detector acts as a spatial integrator, converting the optical intensity distribution into a one-dimensional electrical current. The resulting heterodyne beat frequency component of these two terms is preserved, forming a measurable electrical signal $i^i(t)$,
\begin{equation}
\label{waicha}
i^{i}(t) \propto \frac{1}{2}  \int_{-\frac{D_\text{det}}{2}}^{\frac{D_\text{det}}{2}} \left[ E_d^{i}(t, \boldsymbol{\rho}_d)  \cdot E_{\mathrm{lo}}^*(t) + c.c. \right]d\boldsymbol{\rho}_d,
\end{equation}

where $D_\text{det}$ denotes the diameter of the balance detector aperture. After removing the direct current (DC) component and high-frequency carrier terms, the measurable electrical signal $i^i(t)$ can be simplified as:
\begin{equation}
\label{IF_current}
i^{i}(t) \propto   \exp(j\phi_i) \cdot \iint \tilde{H}\left(\rho_{d}, \rho_{o}\right) P^{i}\left(\rho_{o}\right) \widetilde{T}(\rho_{o})d\boldsymbol{\rho}_dd\boldsymbol{\rho}_o,
\end{equation}

where $\exp(j\phi_i)$ denotes the phase perturbation generated by atmospheric turbulence and noise within the $i$th pulse, in case the transmission medium remains relatively stable within a single detection pulse, all the phase terms caused by the transmission are simplified as a non-temporal variable $\tilde{H}\left(\rho_{d}, \rho_{o}\right)$, including the transmitting quadratic phase term and the transmission function between the target surface and the detector surface. Introducing the vibrating term into Eq. \ref{IF_current}:
\begin{equation}
\begin{aligned}
\label{vibration}
&i^{i}(t) \propto \exp \left(j \phi_{i}\right) \iint \tilde{H}\left(\rho_{d}, \rho_{o}\right) P^{i}\left(\rho_{o}\right) \widetilde{T}\left(\rho_{o}\right) \\
&\times \exp \left[\frac{4 j \pi A\left(\rho_{o}\right)}{\lambda} \cos \left(2 \pi f_{v} t+\phi_{0}(\rho_{o})\right)+j \varphi\left(\rho_{o}\right)\right] d \rho_{d} d \rho_{o}.
\end{aligned}
\end{equation}

The physical representations of the amplitude $A(\rho_o)$, initial phase $\phi_0(\rho_o)$, and roughness function $\varphi(\rho_o)$ in Eq. \ref {vibration} are illustrated in Fig. \ref{target_surface}. Crucially, the target's micro-dynamics are embedded in the phase term. By introducing the vibration model, the instantaneous phase modulation is given by:
\begin{equation}
\label{target_phase}
\Phi(\rho_o, t) = \frac{4\pi}{\lambda} A(\rho_o) \cos(2\pi f_v t + \phi_0(\rho_o)) + \varphi(\rho_o)
\end{equation}
This formulation explicitly shows that the temporal vibration signature is coupled with the spatial speckle field in the bucket detector's output, posing an inverse problem for signal reconstruction. Further signal processes are based on this signal mode.

\subsection{Frequency channel self-calibrated algorithm}
In our previous work\cite{liuMicrovibrationModesReconstruction2022}, we assumed an ideal scenario where the sampling frequency was perfectly synchronized with the target's vibration frequency, ensuring identical motion states across all pulses. Under this condition, we coherently integrated signals from each pulse, yielding a high-SNR vibration signal denoted as $i_{sum}(t)$. Subsequent time-frequency analysis of this signal revealed distinct energy concentrations in the time-frequency spectrum, whose positional indices were recorded. Returning to the $i$-th pulse signal $i^i(t)$, we extracted energy values at these indexed positions, forming an energy sequence $s^i(t_n)$ for each pulse over identical time windows whose central time is $t_n$. Within the same time window, the target can be considered quasi-stationary, enabling image reconstruction via correlation-based ghost-imaging algorithms. However, in practical detection scenarios, targets are predominantly non-cooperative. Without prior knowledge of the precise vibration frequency, temporal calibration of the acquired signals is essential to restore the ‘ideal’ synchronized state assumed in our earlier mode.\\

Considering a general case that each pulse contains $M$ sampling points, where $m$ denotes the sampling index with $m=0,1,...,M-1$, and the sampling interval is $\tau$ , we discretize the continuous signal in Eq. \ref{vibration} into a sampled sequence:
\begin{equation}
\begin{aligned}
\label{vibration_discrete}
&i^{i}[m] \propto \exp \left(j \phi_{i}\right) \iint \tilde{H}\left(\rho_{d}, \rho_{o}\right) P^{i}\left(\rho_{o}\right) \widetilde{T}\left(\rho_{o}\right) \\
&\times \exp \left[\frac{4 j \pi A\left(\rho_{o}\right)}{\lambda} \cos \left(2 \pi f_{v} m\tau+\phi_{0}(\rho_{o})\right)+j \varphi\left(\rho_{o}\right)\right] d \rho_{d} d \rho_{o}.
\end{aligned}
\end{equation}

\begin{figure*}[t!]
\centering
\includegraphics[width=1.8\columnwidth,height=0.46\linewidth]{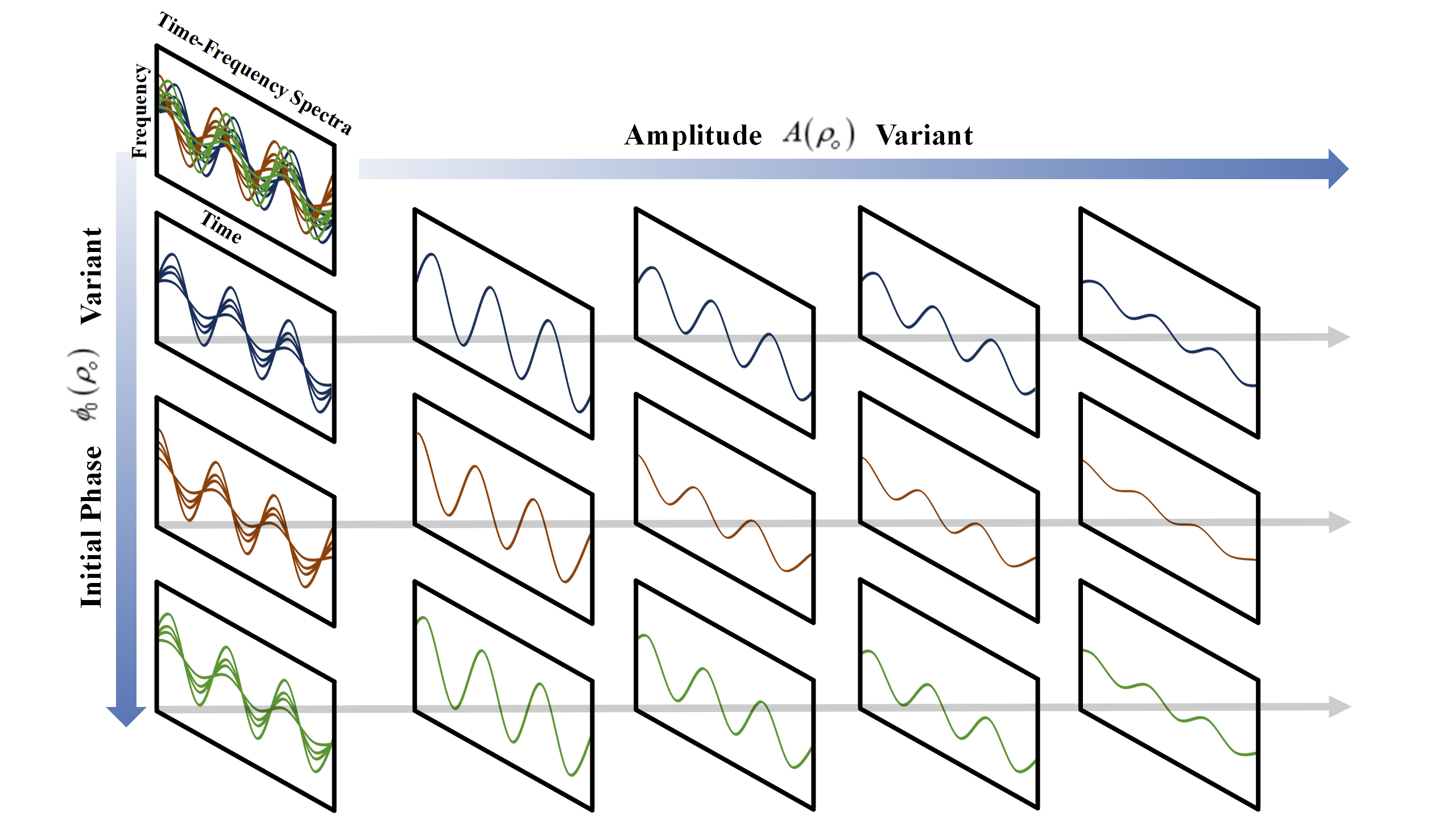}
\caption{A magnified view of a specific probing area on the target surface during detection. The initial phase distribution and amplitude profile across the target surface collectively constitute the vibration mode morphology of the target. On this time-frequency spectrum, the overlapping energy trajectories contributed by scattering points with varying amplitudes and initial phases across the target surface are displayed simultaneously. Similarly, this composite spectrum can be decomposed into multiple distinct, individual time-frequency energy trajectories based on their specific amplitude and initial phase characteristics.}
\label{target_surface}
\end{figure*}

The calibration algorithm first calculates the time offsets derived from the estimated frequency. Subsequently, the raw pulses are temporally aligned based on these offsets to achieve coherent integration. For a random vibrating frequency $f_v$ in a sampling system with a known pulse repetition frequency $f_p$, the number of offset sampling points can be calculated as: 
\begin{equation}
\label{n_offset}
t_{\text {offset }}^{i}=T_{v} \cdot\left(1-\left\{\frac{(i-1) f_{v}}{f_{p}}\right\}\right), \text { if }\left\{\frac{(i-1) f_{v}}{f_{p}}\right\} \neq 0,
\end{equation}
where $T_{v}$ is the vibration period corresponding to the frequency $f_{v}$ and $\{\cdot\}$ denotes the fractional part. Offset time $t_{\text {offet }}^{i}$ can be discretized into sampling points $n_{\text {offet }}^{i}$ which can be expressed as:
\begin{equation}
\label{n_offset}
n_{\text {offset }}^{i}=\operatorname{round}\left(t_{\text {offset }}^{i} \cdot f_{s}\right).
\end{equation}  
Each vibration period comprises $M_v$ sampling points. A fundamental requirement of this algorithm is $M\geq M_v$, ensuring at least one full vibration period per pulse. Under this condition, each pulse encompasses $K = \left\lfloor M / M_{v}\right\rfloor$ complete vibration period, and the sampled sequence of the $k$th segment in the $i$th pulse can be expressed as:
\begin{equation}
\label{segment}
i_{k}^{i}\left[m_{v}\right]=i^{i}\left[m_{v}+k M_{v}\right],
\end{equation} 
where $m_v$ is the sampling index in the vibration with $m_v=0,1,...,M_v-1$. The subsequent signal processing consists of two distinct stages:
\begin{itemize}
    \item \textbf{Alignment}: The $i$th pulse is temporally calibrated by compensating its sampling points offset with reference to the vibration phase at the initial moment of the first pulse. Thus, the aligned $k$th segment $i_{k}^{i, \text { aligned }}\left[m_{v}\right]$ can be written as:
    \begin{equation}
    \label{alignment}
    i_{k}^{i, \text { aligned }}\left[m_{v}\right]=i_{k}^{i}\left[\left(m_{v}+n_{\text {offset }}^{i}\right) \bmod M_{v}\right].
    \end{equation} 
    \item \textbf{Coherent Integration}: Once the temporal alignment is completed according to Eq. \ref{alignment}, the aligned segments from all $N$ pulses and $K$ periods are coherently superimposed. This summation process is critical: it ensures that the signal components matching the calibrated frequency $f_{cal}$ interfere constructively, thereby significantly enhancing the Signal-to-Noise Ratio (SNR), while asynchronous noise and clutter are suppressed. The resulting high-SNR integrated signal, denoted as $i_{sum}[m_v]$, is formulated as:
    \begin{equation}
    \label{integration}
    i_{\text {sum }}\left[m_{v}\right]=\sum_{i=1}^{N} \sum_{k=0}^{K-1} i_{k}^{i, \text { aligned }}\left[m_{v}\right].
    \end{equation} 
\end{itemize}
To visualize the vibration characteristics, Time-Frequency Analysis (TFA) is subsequently applied to the enhanced signal $i_{sum}[m_v]$. In this work, we adopt the Short-Time Fourier Transform (STFT) to generate the energy spectrum. The discrete STFT expression is given by:
\begin{equation}
    \label{STFT}
    E_{spec}[g,h]=S T F T[g,h]=\sum_{l=0}^{L-1} i_{\text {sum }}[l+g S] w[l] e^{-j 2 \pi h l/ N_{f}}.
\end{equation} 
In Eq. \ref{STFT}, $g$ and $h$ are the time window index and frequency window index, respectively, with $g=0,1,...,G-1$ and $h=0,1,...,H-1$. The maximum time window is $G=\left\lfloor\frac{M_{v}-L}{S}\right\rfloor+1$, and $S$ is the length of the sampling points in a single time window. The central time corresponding to each window is $t_g=gS/f_s$. The maximum frequency window is $H$, and $N_f$ is the number of sampled points involved in the fast Fourier transform (FFT). The central frequency of each window is $f_h=hf_s/N_f$. To locate the time/frequency indices of energy concentrations corresponding to the vibrating target in the time-frequency spectrum, the target phase is rewritten from Eq. \ref {vibration_discrete} as the following phase function :
\begin{equation}
    \label{phase function}
    \phi\left(\rho_{o}, m_v\right)=2 k A\left(\rho_{o}\right) \cos \left(2 \pi f_{v} m_v\tau+\phi_{0}(\rho_{o})\right)+\varphi\left(\rho_{o}\right),
\end{equation} 
where $k=2\pi/\lambda$ is the wave vector. Thus, the instant frequency is:
\begin{equation}
\begin{aligned}
    \label{instant frequency}
    f_{\text {instant } }\left(\rho_{o}, m_{v}\right)&=\frac{1}{2 \pi} \frac{d \phi\left(\rho_{o}, m_{v}\right)}{d m_v\tau}\\
    &=-2 k f_{v} A\left(\rho_{o}\right) \sin \left(2 \pi f_{v} m_{v} \tau+\phi_{0}(\rho_{o})\right).
\end{aligned}  
\end{equation}
 In the time-frequency spectrum, the signal energy is concentrated around the instantaneous frequency. For each time frame $g$, the spectral peak occurs at the frequency:
\begin{equation}
    \label{pred}
    \begin{aligned}
   f_{\text {pred }}\left(\rho_{o}, g\right)&=f_{\text {instant }}\left(\rho_{o}, t_{g}\right)\\
   &=-2 k A\left(\rho_{o}\right) \sin \left(2 \pi f_{v} \frac{g S}{f_{s}}+\phi_{0}(\rho_{o})\right).
   \end{aligned}  
\end{equation}
Considering the limit of sampling rate $f_s$, the potential negative frequency is mapped to the positive frequency spectrum:
\begin{equation}
    \label{map}
    f_{\text {map }}(\rho_{o},g)=\left\{\begin{array}{c}
f_{\text {pred }}\left(\rho_{o}, g\right), \text { if } f_{\text {pred }}\left(\rho_{o}, g\right) \geq 0 \\
f_{\text {pred }}\left(\rho_{o}, g\right)+f_{s}, \text { if } f_{\text {pred }}\left(\rho_{o}, g\right)<0.
\end{array}\right.
\end{equation} 
Subsequently, the indices of the target energy in the frequency window are given by:
\begin{equation}
    \label{freq indice}
    f_{\text {peak }}(\rho_{o},g)=\operatorname{round}\left(\frac{N_{f}}{f_{s}} f_{\text {map }}(\rho_{o},g)\right).
\end{equation} 
Due to the target spatial distribution $\rho_o$, the energy in the time-frequency spectrum undergoes broadening as a result of this spatial variation. Returning to the Eq. \ref{instant frequency}, two non-ideal terms are introducing the spatial distribution interference, $A(\rho_o)$ and $\phi_{0}(\rho_{o})$. The former exhibits a spatially dependent amplitude distribution. Consequently, the corresponding time-frequency spectrum no longer appears as a single trajectory, but rather manifests as a frequency band, which can be expressed as:
\begin{equation}
    \label{freq band, amplitude}
    \Delta f_{\text {band,amplitude }}=2 k f_{v} \cdot\left[\max \left(A\left(\rho_{o}\right)\right)-\min \left(A\left(\rho_{o}\right)\right)\right]
\end{equation} 
The latter effect arises from the spatial variation in the initial phase among different target points, leading to phase heterogeneity in the vibration profile. This spatial distribution of initial phases introduces asynchrony across the target surface, thereby causing the time-frequency spectrum to transition from a distinct trajectory to a broadened frequency band. For the same amplitude $A_s$, the energy occurs in,
\begin{equation}
    \begin{aligned}
    \label{freq band, phase}
   \Delta f_{\text {band, phase }}&=2 k f_{v} \cdot[\max _{\rho_{o}}(-2 k A_{s} f_{v} \sin (2 \pi f_{v} t_{g}+\phi_{0}(\rho_{o})))\\
   &-\min _{\rho_{o}}(-2 k A_{s} f_{v} \sin (2 \pi f_{v} t_{g}+\phi_{0}(\rho_{o})))].
 \end{aligned}
 \end{equation}
While this transformation increases the complexity of target detection and feature extraction, it simultaneously provides additional information regarding the spatial structure of the target. Thus, the initial phase scan of the entire target can reconstruct the structure of the target surface.





\begin{algorithm}[!t]
\caption{Frequency-Channel Self-Calibration}
\label{suanfa1}
\begin{algorithmic}[1]
\renewcommand{\algorithmicrequire}{\textbf{Input:}}
\renewcommand{\algorithmicensure}{\textbf{Output:}}

\REQUIRE Discretized echo sequence $i^i[m]$ (Eq. \ref{vibration_discrete}); \\
Calibration frequency channel $f_{v}$.

\ENSURE Calibrated Energy Spectrum $\mathbf{E}_{spec}$.

\vspace{0.1cm}
\STATE \textbf{Step 1: Temporal Phase Alignment (Virtual Synchronization)}
\FOR{each pulse $i = 1$ to $N$}
    \STATE Calculate temporal offset $t_{offset}^i$ based on $f_{v}$ (Eq. \ref{n_offset}).
    \STATE Apply cyclic shift to segment $k$ to obtain $i_k^{i, \text{aligned}}[m_v]$ (Eq. \ref{segment}, Eq. \ref{alignment}).
\ENDFOR

\vspace{0.1cm}
\STATE \textbf{Step 2: Coherent Integration}
\STATE Sum aligned segments to enhance SNR (Eq. \ref{integration}):
\STATE \quad $i_{sum}[m_v] \leftarrow \sum_{i=1}^{N} \sum_{k=0}^{K-1} i_k^{i, \text{aligned}}[m_v]$

\vspace{0.1cm}
\STATE \textbf{Step 3: Time-Frequency Analysis}
\STATE Compute STFT of the integrated signal $i_{sum}[m_v]$ (Eq. \ref{STFT}).
\STATE Calculate Magnitude Spectrum: $\mathbf{E}_{spec}[g, h] \leftarrow |\text{STFT}[g, h]|^2$.

\RETURN $\mathbf{E}_{spec}$
\end{algorithmic}
\end{algorithm}

\subsection{Dominant Scatterer Micro-Doppler Tracking and Microvibration Mode Reconstruction}
After the calibrated time-frequency spectrum is acquired, the primary task is to identify the energy sequence of the dominant scatterer for subsequent use in GI-based reconstruction algorithms. The complete procedure of frequency calibration and energy trajectory extraction are shown in the Alg.\ref{suanfa1} In this work, Particle Swarm Optimization (PSO) is employed to implement the microvibration tracking process. The algorithm searches for the optimal vibration parameters that best fit the micro-Doppler trajectory of the dominant scatterer. PSO is a stochastic optimization technique grounded in swarm intelligence, mimicking the social information-sharing mechanism observed in bird flocking behavior\cite{kennedyParticleSwarmOptimization1995}. In this application, each particle represents a candidate position in the time-frequency domain, specifically, a combination of the time window index $g$ and the frequency index $h$. The particle swarm iteratively explores the solution space to locate the global maximum of energy. During the initialization, the positions and velocities of the particles are randomly assigned. In each iteration, every particle is found by the entire swarm. A fitness function is defined to evaluate the quality of each particle's position. Through repeated updates, the swarm gradually converges toward the region in the time-frequency plane with the highest energy concentration, allowing for the accurate extraction of the target's scattering characteristics. Defining the optimal parameter vector $\textbf{p}^*$ as

\begin{figure}[htbp]
\centering
\includegraphics[width=0.97\columnwidth,height=1.2\linewidth]{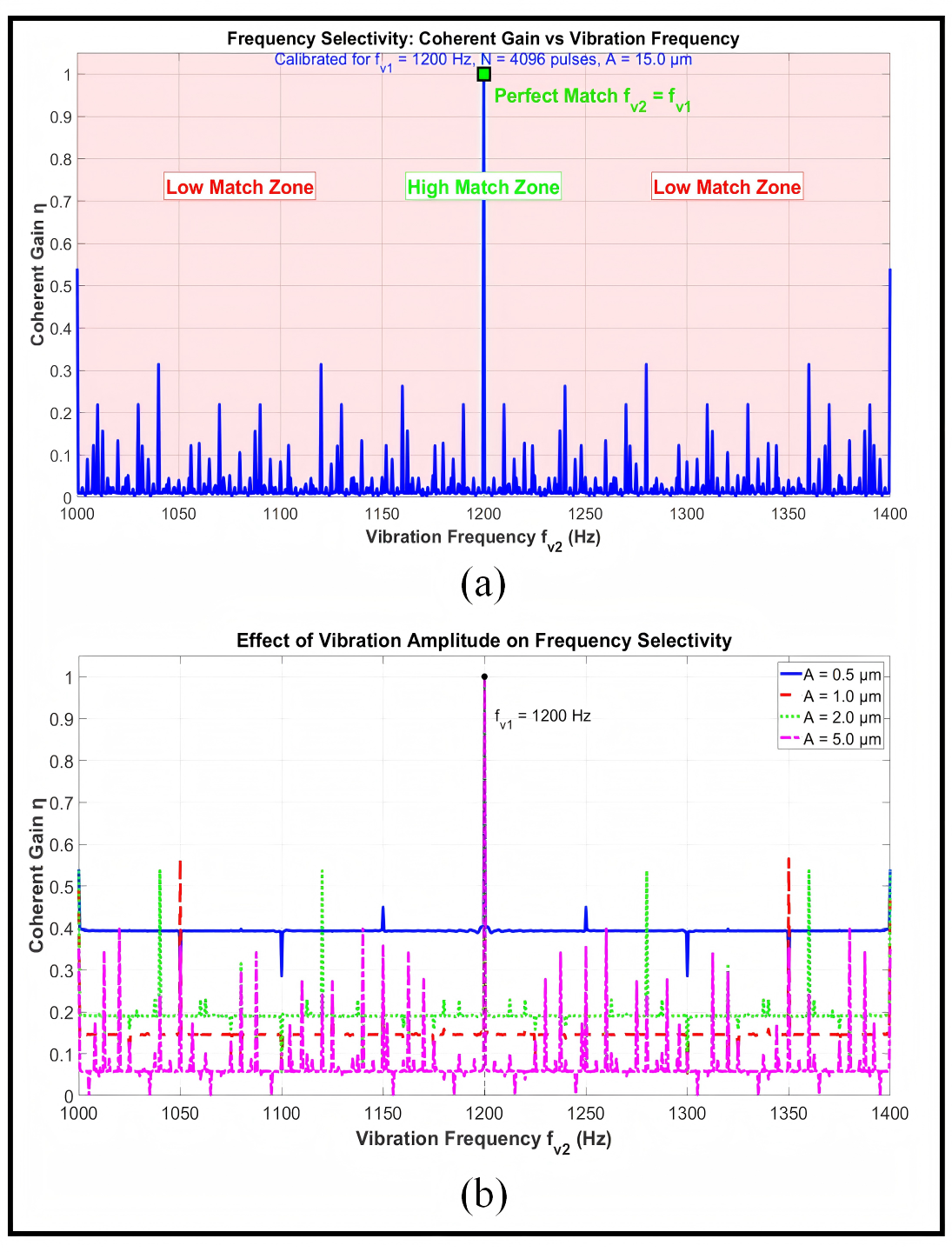}
\caption{Simulation result related to the performance of the frequency selectivity. (a) displays the coherent gain curves of two point targets when $f_{v1}$ is set to 1200Hz and $f_{v2}$ is varied. The green region indicates the 3dB bandwidth, which corresponds to the high match zone. The red region represents the low match zone. (b) illustrates how the coherent gain curve evolves under different vibration amplitudes. It is clearly visible that as the modulation depth of the Bessel function increases, the coherence degrades more rapidly with oscillations, as expected from the theoretical model.}
\label{simulation_chart}
\end{figure}

\begin{equation}
    \label{dictionary}
    p^*=(A^*(\rho_o^{\max}),f^*_v,\phi{o}^*(\rho_0^{\max})),
\end{equation} 
where $\rho_o^{max}$ is the target region providing the dominant scatterer energy. Therefore, the optimal trajectory can be expressed as:
\begin{equation}
\begin{aligned}
    \label{trajectory}
    \mathrm{P}^{*}=&\{(g, f^{*}(g)) \lvert\, f^{*}(g)=\operatorname{round}(-\frac{F}{f_{s}} A^{*}\left(\rho_{o}^{\max }\right)\\ &\times\sin \left(2 \pi f_{v}^{*} t_{g}+\phi_{0}\left(\rho_{o}^{\max }\right)\right))., t_{g}=\frac{g S}{f_{s}}\}.
\end{aligned}
\end{equation}
Combining this with Eq. \ref{vibration_discrete}, the energy summation along the optimized trajectory effectively smooths the vibrational term $sin(\bullet)$. Therefore, the total energy $E^i$ along the optimal trajectory for the $i$th pulse can be written as:
\begin{equation}
\label{ith energy}
E^i\propto\left|A_{LO}\int d\rho_d\int d\rho_o^{\max}\exp\left[j\varphi_{LO}\left(\rho_o^{\max},\rho_d\right)\right]T\left(\rho_o^{\max}\right)\right|^2.
\end{equation}
After acquiring the whole energy sequence of the interesting target area $T(\rho_o^{max})$, the corresponding target imaging can be reconstructed by applying second-order correlation, a core principle of GI, as expressed by the following process:
\begin{equation}\begin{aligned}
\label{final image}
  G^2\left(\rho_o^{\max}\right)&=\left\langle P^i\left(\rho_o\right)\cdot E^i\right\rangle-\left\langle P^i\left(\rho_o\right)\right\rangle\cdot\left\langle E^i\right\rangle \\
 & \propto T\left(\rho_o^{\max}\right).
\end{aligned}\end{equation}

\begin{figure}[htbp]
\centering
\includegraphics[width=0.97\columnwidth,height=0.75\linewidth]{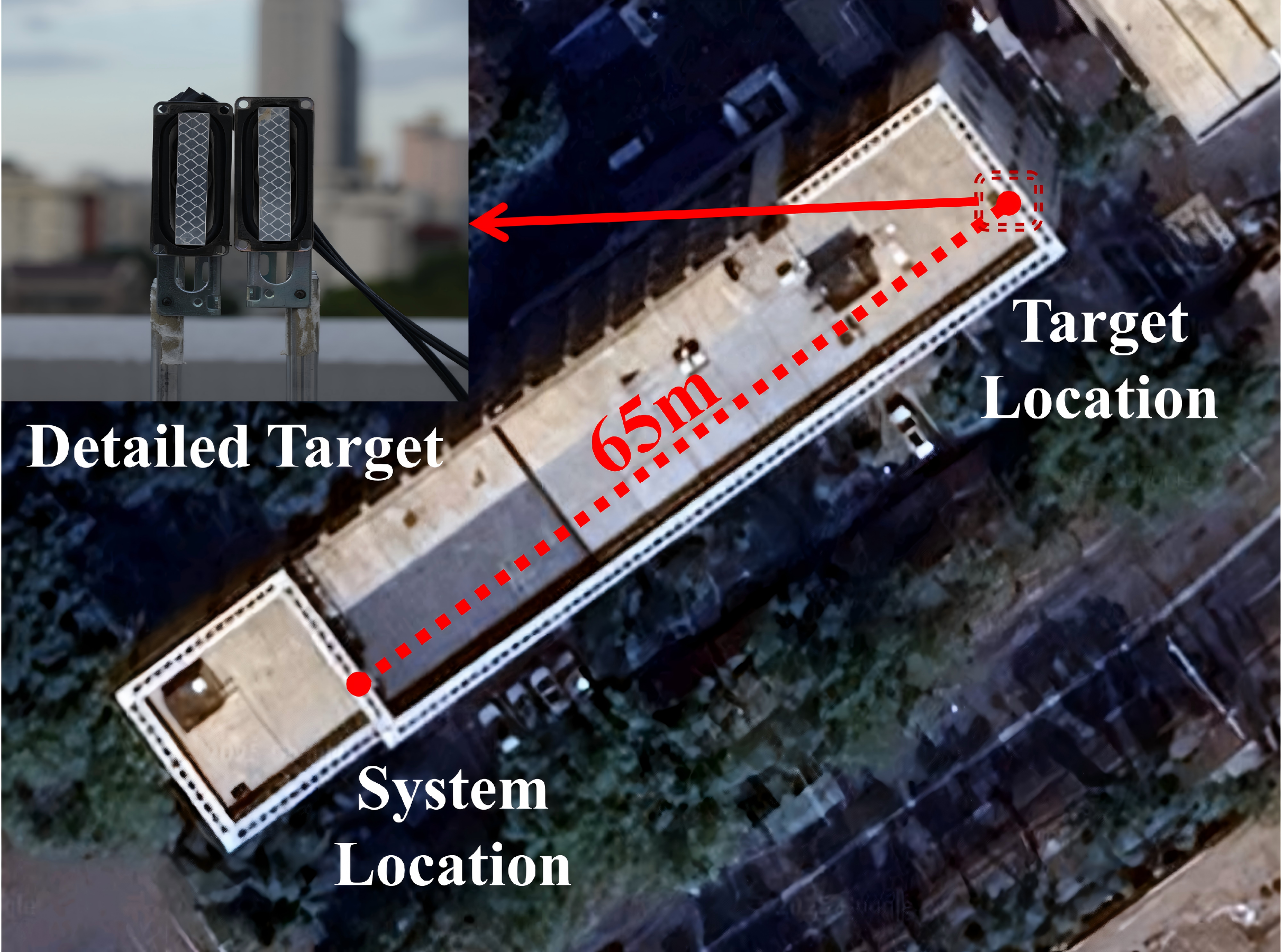}
\caption{A satellite view of the experimental site. The top-left corner shows a detailed image of the target area.}
    \label{Exp_config}
\end{figure}



\begin{algorithm}[!t]
\caption{Dominant Scatterer Micro-Doppler Tracking and Dynamic Reconstruction}
\label{alg2}
\begin{algorithmic}[1]
\renewcommand{\algorithmicrequire}{\textbf{Input:}}
\renewcommand{\algorithmicensure}{\textbf{Output:}}

\REQUIRE Calibrated Spectrum $\mathbf{E}_{spec}$ (from Alg. \ref{suanfa1}); \\
Reference patterns $P^i(\rho_o)$ (DMD patterns); \\
Search space $\mathcal{S} = [A, f_v, \phi_0]$.

\ENSURE Spatio-temporal vibration video frames $\mathbf{G}_{video}$.

\vspace{0.1cm}
\STATE \textbf{Stage 1: Micro-Doppler Tracking (via PSO)}
\STATE Initialize particles.
\WHILE{not converged}
    \STATE Calculate theoretical trajectory $f_{track}(g)$ for each particle (Eq. \ref{trajectory}).
    \STATE Evaluate Fitness: Sum of $\mathbf{E}_{spec}$ energy along $f_{track}(g)$.
    \STATE Update particle states to maximize Fitness.
\ENDWHILE
\STATE \textbf{Obtain} Optimal Parameters $\mathbf{p}^* = [A^*, f_v^*, \phi_0^*]$.

\vspace{0.1cm}
\STATE \textbf{Stage 2: Dynamic Reconstruction (Phase Scanning)}
\FOR{scan phase $\alpha \in [0, 2\pi]$}
    \STATE \textit{1) Virtual Phase Shifting:} Update $\phi_{new} \leftarrow \phi_0^* + \alpha$.
    \STATE \textit{2) Energy Extraction:} Integrate energy $E^i(\alpha)$ along the shifted trajectory defined by $[A^*, f_v^*, \phi_{new}]$ (Eq. \ref{ith energy}).
    \STATE \textit{3) Imaging:} Reconstruct frame $I_{\alpha}$ via Second-Order Correlation (Eq. \ref{final image}).
    \STATE Append $I_{\alpha}$ to $\mathbf{G}_{video}$.
\ENDFOR

\RETURN $\mathbf{G}_{video}$
\end{algorithmic}
\end{algorithm}

Consequently, the image of the dominant scatterer is initially reconstructed as in Eq. \ref{final image}. However, this equation yields only a static snapshot corresponding to the fixed initial phase $\phi_0^*$ found by PSO. To visualize the full temporal evolution of the vibration mode, we implement a virtual phase scanning strategy based on the identified optimal parameters $\mathbf{p}^*=[A^*, f_v^*, \phi_0^*]$.Instead of using a fixed phase, we systematically shift the phase term in the reference trajectory model, which is explained in Eq. \ref{trajectory} across the range $[0, 2\pi]$. For each discrete phase step $\alpha$, the trajectory is virtually displaced in the time-frequency domain, and a new reference energy vector is integrated along this shifted path. We define the ghost image reconstructed from this specific phase step as a single video frame, denoted as $I_\alpha$. The sequential assembly of these frames constitutes the complete dynamic vibration video, formally expressed as:
\begin{equation}
\label{fv1_phase}
\mathbf{G}_{video} = \{I_\alpha \mid \alpha \in [0, 2\pi]\}.
\end{equation}
This process is detailed in Alg. \ref{alg2}, allows for the frame-by-frame visualization of the target's dynamic microvibration.

\begin{figure*}[htbp]
\centering
\includegraphics[width=2\columnwidth,height=0.2\linewidth]{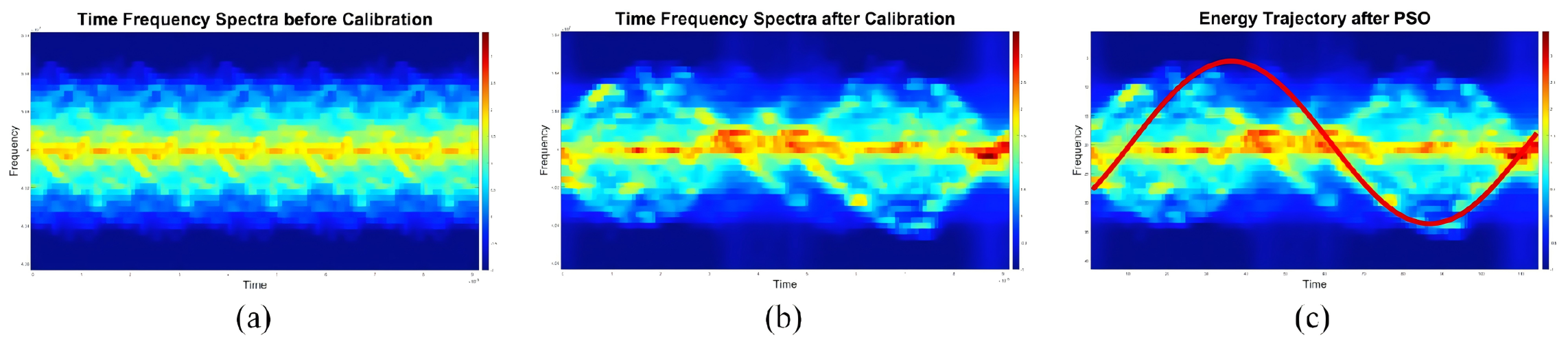}
\caption{Visualization of the time-frequency spectrum after calibration/PSO processing. $(a)$ shows the result of direct STFT-based coherent integration applied to the uncalibrated echo signal. $(b)$ presents the output of the coherent integration after frequency calibration. $(c)$ displays the optimized energy trajectory with the target energy weighted obtained by PSO, along with the corresponding estimated amplitude and initial phase values. The red trajectory shows the optimized energy trajectory.}
    \label{TF_spectrum}
\end{figure*}

\section{Performance of the Frequency Selectivity}
In realistic detection scenarios, the FOV may contain multiple targets characterized by distinct vibration frequencies. By aligning signals within each frequency channel, coherent integration can be realized during imaging, thereby enabling selective extraction of specific coherent frequency components. This section presents a theoretically derived performance mode for the frequency calibration algorithm used to screen different frequency channels.\\
Assuming two targets are present within the FOV, with the frequencies $f_{v1}$ and $f_{v2}$. When the signal is perfectly calibrated according to the frequency $f_{v1}$, the initial vibration phase of the target oscillating at this frequency at the start of the $i$th pulse is given by:
\begin{equation}
\label{fv1_phase}
\Phi_{f_{v 1}}^{i}\left(\rho_{o}\right) = 2 \pi f_{v 1} t_{i}+\phi_{01}\left(\rho_{o}\right), t_{i} = \frac{i-1}{f_{p}}.
\end{equation}
Consequently, the phase offset of the $i$th pulse is given by:
\begin{equation}
\label{fv1_phase}
\delta_{i}=2 \pi f_{v 1} \cdot \frac{i-1}{f_{p}} \bmod 2 \pi.
\end{equation}
Then the cyclic shift offset sampling points $n_{\text {offset}}^{i}$ can be written as:
\begin{equation}
\label{fv1_n_offset}
n_{\text {offset}}^{i}=\operatorname{round}\left(\frac{\delta_{i}}{2 \pi} M_{v 1}\right).
\end{equation}  
The equivalent time offset $\tau_i$, corresponding to a cyclic shift offset of $n_{\text {offset}}^{i}$ is calculated as:
\begin{equation}
\label{equivalent time offset}
\tau_{i} \approx \frac{1}{f_{v 1}}\left(f_{v 1} \frac{i-1}{f_{p}} \bmod 1\right).
\end{equation}  
Following the calibration, the initial vibration phase of the target with the vibration frequency $f_{v2}$ is updated to:
\begin{equation}
\label{fv2_offset_phase}
\begin{array}{l}
\Phi_{f_{v 2}, \text { offset }}^{i}\left(\rho_{o}\right)=2 \pi f_{v 2}\left(t_{i}-\tau_{i}\right)+\phi_{02}\left(\rho_{o}\right) \\
=\Phi_{f_{v 2}}^{i}\left(\rho_{o}\right)+2 \pi f_{v 2} \frac{i-1}{f_{p}}-2 \pi \frac{f_{v 2}}{f_{v 1}}\left(f_{v 1} \frac{i-1}{f_{p}} \bmod 1\right) \\
=\Phi_{f_{v 2}}^{i}\left(\rho_{o}\right)+2 \pi \frac{f_{v 2}}{f_{v 1}}\left\lfloor f_{v 1} \frac{i-1}{f_{p}}\right\rfloor,
\end{array}
\end{equation}

\begin{figure}[htbp]
\centering
\includegraphics[width=0.97\columnwidth,height=1.0\linewidth]{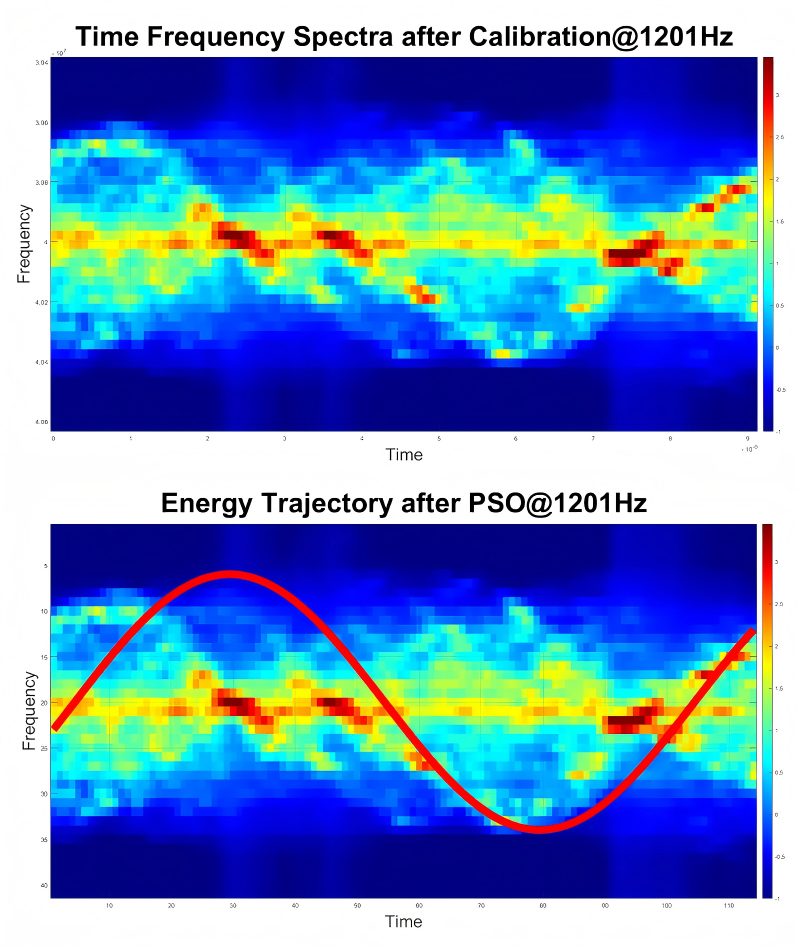}
\caption{Results of the 1201 Hz calibrated frequency channel, obtained using the same methodology shown in Fig.\ref{TF_spectrum}.}
    \label{1201_spectrum}
\end{figure}
where $\Phi_{f_{v2}}^{i}\left(\rho_{o}\right)$ is the original initial vibration phase of the target with the vibration frequency $f_{v2}$. For the vibration frequency $f_{v2}$, the coherently integrated signal is given by:

\begin{equation}
\label{sum}S_{f_{v2}}
\propto\sum_{i=0}^N\int d\rho_o\exp\left(j2kA(\rho_o)\sin\left[\Phi_{f_{v2},\text { offset }}^i\left(\rho_o\right)\right]\right).
\end{equation}

The analysis of an extended target proves overly complex. To more intuitively demonstrate the frequency resolution achieved after coherent integration, we simplify the scenario by reducing the extended target to a single-point target. For the coherently integrated signal $S_{f_{v2}}$, the expression simplifies to:
\begin{equation}
\label{simplify_sum}
S_{f_{v2}}\propto\sum_{i=0}^N\exp\left(j2kA\sin\left(\Phi_{f_{v2},\text { offset }}^i\right)\right).
\end{equation}

When the target model degenerates to a single point target, the Jacobi-Anger expansion can be applied as:
\begin{equation}S_{f_{v2}}
\label{Jacob-anger}\propto\sum_{i=0}^{N-1}\sum_{n=-\infty}^\infty j^nJ_n\left(2kA\right)\exp\left(j2n\pi f_{v2}\left(t_i-\tau_i\right)+\phi_{02}\right),\end{equation}
where, $J_n(\cdot)$ denotes the $n$th Bessel function. Consequently, the final normalized coherent gain $\eta$ can be expressed as:
\begin{equation}
\label{normalized eta}
\eta=\frac{\left|S_{f_{v 2}}\right|}{N}.
\end{equation}

Simulation results of correlation performance under the experimental parameters are presented in Fig. \ref{simulation_chart}. For clarity, the simulation considers two-point monochromatic targets. A total of 4096 pulses are transmitted. The reference target vibrates at frequency $f_{v1}$, while the vibration frequency $f_{v2}$ of the second target is varied from 1000Hz to 1400Hz. Fig. \ref{simulation_chart}.$(a)$ shows the correlation trajectories between two targets at different frequency configurations when both have an amplitude of $15\mu m$. Fig. \ref{simulation_chart}.$(b)$ illustrates the influence of amplitude variation on the coherent gain.

\begin{figure*}[htbp]
\centering
\includegraphics[width=2\columnwidth,height=0.29\linewidth]{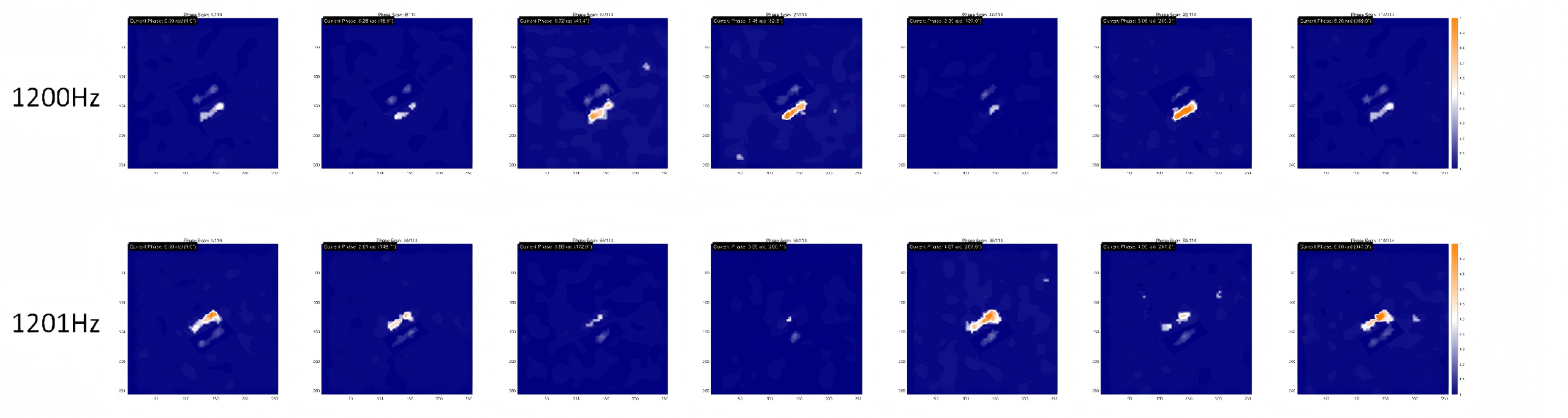}
\caption{The figure displays imaging results for different initial phase values. Images in the first row correspond to the 1200 Hz frequency channel, while those in the second row are from the 1201 Hz channel.}
    \label{1200/1201_special_ponits}
\end{figure*}

\section{Experiment Result}
To rigorously validate the reliability of the proposed algorithm and the underlying principles, we conducted an experimental validation using a coherent laser ghost imaging radar system. As shown in Fig. \ref{Exp_config}, the experiment was performed on the rooftop of a building. The target is located 65 meters from the system. The target consists of two loudspeakers independently driven by separate audio sources, whose surfaces are covered with a high-reflectivity film to enhance their reflectivity.

A computer generated two separate single-frequency signals at 1200 Hz and 1201 Hz. The two signals were fed to two separate loudspeakers. The CD-GI lidar system was then employed to detect these vibrating targets. A balanced detector is used to acquire the one-dimensional electrical signal containing spatiotemporal information of the target vibration mode.\\
The received one-dimensional electrical signal is processed through the workflow illustrated in Fig. \ref{TF_spectrum}. The purpose of calibration is to enable coherent integration, accumulating the energy signal in the corresponding frequency channels. Fig.\ref{TF_spectrum} demonstrates calibration based on 1200 Hz, resulting in coherent integration of the target vibrating at this specific frequency. Through PSO, the strongest energy-weighted trajectory is identified, which can be interpreted as the vibration profile of the dominant scattering point within a detecting pulse period.\\

Similarly, the same raw signal can be calibrated at the 1201Hz frequency channel. The resulting time-frequency spectrum and the strongest energy-weighted trajectory of the dominant scatterer after calibration are shown in Fig. \ref{1201_spectrum}.\\

By comparing the time frequency spectra shown in Fig. \ref{TF_spectrum} and Fig. \ref{1201_spectrum}, it can be observed that selecting different frequency channels results in calibrating significantly distinct spectrum energy distributions. The strongest energy-weighted trajectories obtained via PSO also exhibit completely different initial phases. Following the algorithmic workflow, the time windows and the frequency windows indices traversed by the strongest energy-weighted trajectory are acquired, and the energy values at corresponding indices across all pulses are incoherently integrated. The resulting energy sequence then undergoes the second-order correlation. The output of the second-order correlation corresponds to the spatial position of the dominant scatterer at the respective frequency, enabling spatial localization of targets vibrating at different frequencies. Applying this method separately to the dominant scattering points at 1200 Hz and 1201 Hz, the 3-dimensional reconstructed images are shown in Fig. \ref{3D_result}.\\
\begin{figure}[htbp]
\centering
\includegraphics[width=1.1\columnwidth,height=0.7\linewidth]{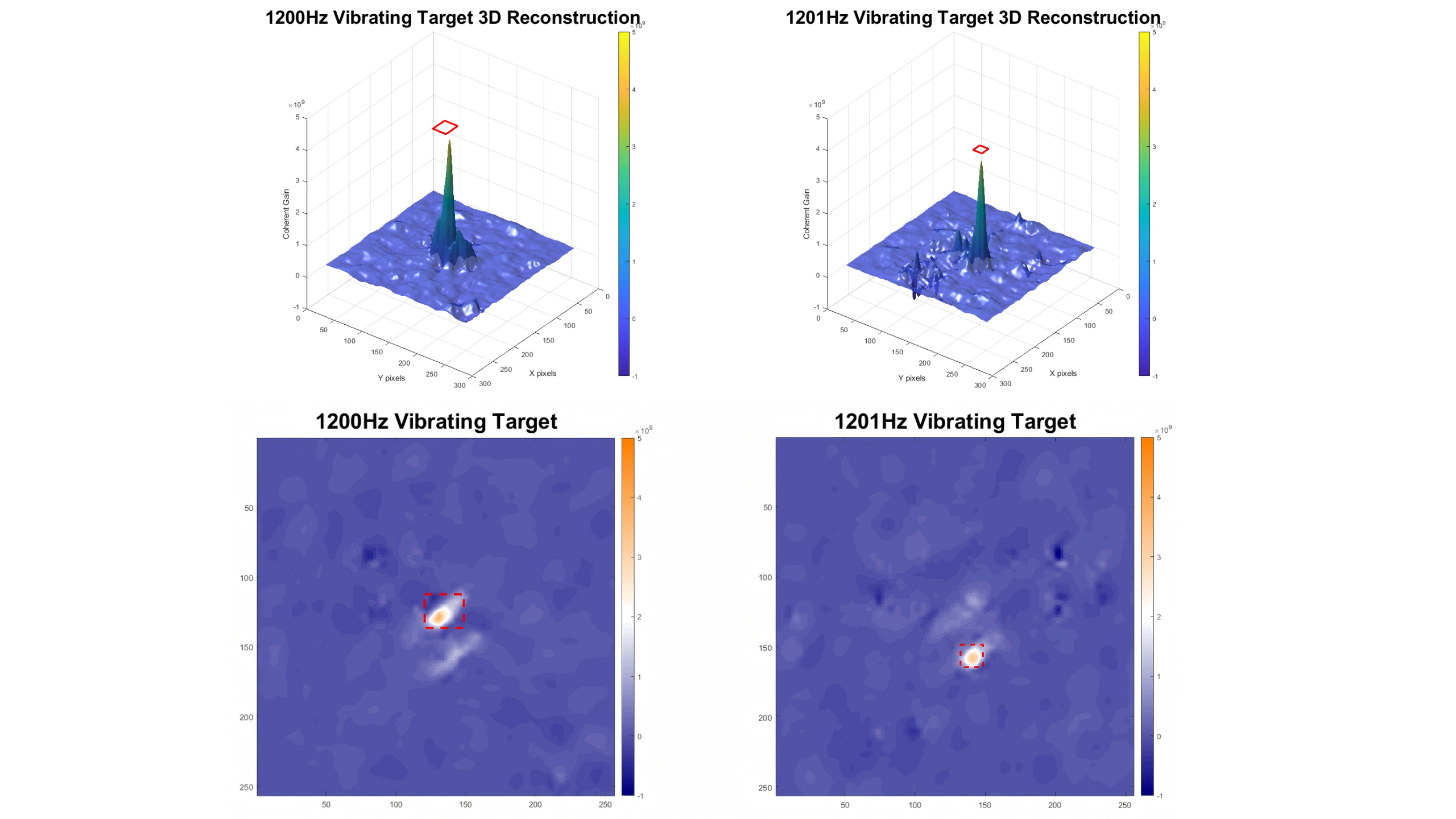}
\caption{The second-order correlation results for the target dominant scatterer, derived from the spectrum calibrated at 1200 Hz and 1201 Hz, respectively. These two images are displayed with the same FOV angles. The first row is the 3-dimensional reconstruction. The second row is the 2-dimensional reconstruction. Red-dashed bounding boxes highlight the scattered points.}
    \label{3D_result}
\end{figure}

After obtaining the time-frequency characteristic trajectory of the dominant scatterer, the optimization algorithm outputs its amplitude, frequency, and initial phase values. As in the theoretical derivation above, a horizontal shift of the entire time-frequency characteristic trajectory can implement the initial phase scan. Keeping the amplitude and frequency of the characteristic trajectory constant, we scan the initial phase over a full cycle from $0$ to $2\pi$ with fixed steps, acquiring an image at each step. Each such image is recorded as one frame; Thus, completing the phase scan yields a full video of the target's vibration mode. As shown in Fig. \ref{1200/1201_special_ponits}, we have selected several representative frames from this vibration mode video for demonstration. Due to the high similarity between the two targets, we applied a mask to suppress responses in non-regions of interest, leveraging the localization capability shown in Fig. \ref{3D_result}. This masking operation can be implemented using morphological detection algorithms \cite{bobinMorphologicalComponentAnalysis2007}. The complete scan video of the target vibration mode is provided in the supplementary materials. As is visually evident, even for a geometrically simple vibrating target like the loudspeaker, the scanning process reveals both a central single-point structure and a separated two-point structure on opposite sides. Disassembly analysis confirms that this specific loudspeaker type indeed contains two piezoelectric vibration elements beneath its leather membrane, which act as the dual excitation sources. This finding directly validates that the detection results correspond to the physical vibration mode of the target, thereby providing supporting evidence for the theoretical framework.

\section{Discussion}
\subsection{Spectral Selectivity and Hyper-Fine Frequency Resolution}
The experimental results presented in Fig. \label{3D_result} demonstrate a remarkable capability of the proposed CD-GI system: the spatial discrimination of targets with a frequency difference as small as 1 Hz. This hyper-fine spectral resolution stems from the inherent frequency selectivity of our frequency-channel self-calibration algorithm.Unlike conventional wideband imaging, our method effectively functions as a tunable narrowband "spatial-temporal filter." By mathematically locking the integration window to a specific frequency channel $f_{cal}$, signals varying at $f_{cal}$ constructively interfere, resulting in a high coherent gain, whereas asynchronous signals (even those differing by only a few Hertz) suffer from destructive interference and are naturally suppressed. This feature is particularly advantageous in complex remote sensing scenarios where multiple vibration sources or mechanical modes coexist. It allows for the independent "slicing" and reconstruction of each vibration mode from a single composite bucket signal, a feat unattainable by standard intensity-based sensing.
\subsection{Physics of Amplitude Modulation and Bessel Decomposition}
While this work primarily focuses on the morphological reconstruction of microvibration modes (i.e., identifying where and how the target vibrates), the underlying signal model offers a clear pathway toward quantitative amplitude measurement. As derived in our theoretical analysis as explained in Eq. \ref{normalized eta}, the spectral structure of the echo signal is governed by the Jacobi-Anger expansion, where the magnitude of harmonic components is directly linked to the vibration amplitude $A(\rho_o)$ via Bessel functions of the first kind, $J_n(4\pi A/\lambda)$. This non-monotonic relationship implies that the "modulation depth" observed in the time-frequency spectrum carries quantitative information about the physical displacement. Although the current algorithm optimizes for the fundamental frequency component to maximize SNR for shape reconstruction, future iterations could exploit the ratio of higher-order harmonics to precisely retrieve the absolute vibration amplitude at the sub-wavelength scale, bridging the gap between qualitative imaging and quantitative metrology.
\subsection{Limitations and Future Scope}
Despite the robust performance demonstrated on non-cooperative monochromatic targets, several challenges remain to be addressed. First, the current "Phase Scanning" and alignment strategy assumes a quasi-stationary vibration spectrum within the detection window. For targets exhibiting rapid transient responses or wideband random vibrations, the single-frequency calibration assumption may need to be generalized to a joint time-frequency analysis framework, such as using Wavelet Transforms instead of STFT.

Second, the reconstruction quality relies on the sparsity of the dominant scatterers. In scenarios with extremely low surface reflectivity or dense, distributed scattering, the "dominant scatterer tracking" approach might face convergence issues. Integrating advanced compressive sensing priors or deep learning-based denoising could further enhance the system's robustness in low-SNR environments.

\section{Conclusion}
This paper presents a novel CD-GI framework capable of resolving the full-field microvibration dynamics of non-cooperative targets. Addressing the fundamental limitations of point-scanning vibrometry and resolution-constrained array sensors, we formulated the micro-vibration imaging task as a mathematical inverse problem involving spatio-temporal coupling.

To solve this, we developed a frequency channel self-calibration scheme combined with a dominant scatterer micro-Doppler tracking algorithm. This methodology successfully decouples the temporal vibration signatures from spatial speckle patterns without requiring prior knowledge of the target's microvibration parameters or external synchronization. Experimental validation confirmed that the system achieves sub-wavelength microvibration sensitivity and is capable of spatially distinguishing adjacent targets with a frequency separation of only 1 Hz.

By synergizing the high sensitivity of coherent detection with the spatial multiplexing efficiency of computational GI, this work provides a robust, scalable solution for structural health monitoring, aerospace inspection, and remote acoustic sensing. Future developments will focus on extending this paradigm to quantitative amplitude metering and broadband vibration analysis, further broadening its application potential in complex dynamic environments.

\bibliographystyle{ieeetr}
\bibliography{Bib}

\begin{thebibliography}{10}

\bibitem{muellerComparingGRACEFOKBR2022}
V.~Mueller, M.~Hauk, M.~Misfeldt, L.~Mueller, H.~Wegener, Y.~Yan, and G.~Heinzel, ``Comparing {GRACE}-{FO} {KBR} and {LRI} {Ranging} {Data} with {Focus} on {Carrier} {Frequency} {Variations},'' {\em Remote Sensing}, vol.~14, p.~4335, Sept. 2022.
\newblock Num Pages: 29 Web of Science ID: WOS:000851914900001.

\bibitem{zhangRemovalMicroDopplerEffect2021}
S.~Zhang, Y.~Liu, X.~Li, and D.~Hu, ``Removal of {Micro}-{Doppler} {Effect} of {ISAR} {Image} {Based} on {Laplacian} {Regularized} {Nonconvex} {Low}-{Rank} {Representation},'' {\em IEEE Transactions on Image Processing}, vol.~30, pp.~6446--6458, 2021.

\bibitem{zhangMicroDopplerEffectsRemoved2021}
S.~Zhang, Y.~Liu, and X.~Li, ``Micro-{Doppler} {Effects} {Removed} {Sparse} {Aperture} {ISAR} {Imaging} via {Low}-{Rank} and {Double} {Sparsity} {Constrained} {ADMM} and {Linearized} {ADMM},'' {\em IEEE Transactions on Image Processing}, vol.~30, pp.~4678--4690, 2021.

\bibitem{chenMicrodopplerEffectRadar2006}
V.~C. Chen, F.~Y. Li, S.~S. Ho, and H.~Wechsler, ``Micro-doppler effect in radar: {Phenomenon}, model, and simulation study,'' {\em Ieee Transactions on Aerospace and Electronic Systems}, vol.~42, pp.~2--21, Jan. 2006.
\newblock Num Pages: 20 Web of Science ID: WOS:000236288000001.

\bibitem{junDetectionMicroDopplerEffect2014}
Z.~Jun, S.~Yang, C.~Zenghui, S.~Tengfei, and Z.~Tiantian, ``Detection on micro-{Doppler} effect based on 1550nm laser coherent radar,'' {\em Infrared Physics \& Technology}, vol.~62, pp.~34--38, Jan. 2014.

\bibitem{lachettaSimulatingDigitalMicromirror2021}
M.~Lachetta, H.~Sandmeyer, A.~Sandmeyer, J.~S.~a. Esch, T.~Huser, and M.~Müller, ``Simulating digital micromirror devices for patterning coherent excitation light in structured illumination microscopy,'' {\em Philosophical Transactions of the Royal Society A: Mathematical, Physical and Engineering Sciences}, vol.~379, p.~20200147, Apr. 2021.

\bibitem{rothbergInternationalReviewLaser2017}
S.~J. Rothberg, M.~S. Allen, P.~Castellini, D.~Di~Maio, J.~J.~J. Dirckx, D.~J. Ewins, B.~J. Halkon, P.~Muyshondt, N.~Paone, T.~Ryan, H.~Steger, E.~P. Tomasini, S.~Vanlanduit, and J.~F. Vignola, ``An international review of laser {Doppler} vibrometry: {Making} light work of vibration measurement,'' {\em Optics and Lasers in Engineering}, vol.~99, pp.~11--22, Dec. 2017.
\newblock Num Pages: 12 Web of Science ID: WOS:000413176100003.

\bibitem{vasiliuBasePlateResonance2024}
A.-F. Vasiliu, E.-S. Csukas, P.-I. Virga, and D.~Comeaga, ``Base {Plate} {Resonance} {Frequencies} {Determination} via a {Laser} {Vibrometer}: {EMA}, {FEA}, and {CrossMAC} {Validation},'' in {\em Advances in 3om: {Opto}-{Mechatronics}, {Opto}-{Mechanics}, and {Optical} {Metrology}, 3om 2023} (V.~F. Duma, J.~P. Rolland, A.~G.~H. Podoleanu, M.~Guina, and C.~Sinescu, eds.), vol.~13187, (Bellingham), p.~131870F, Spie-Int Soc Optical Engineering, 2024.
\newblock Num Pages: 10 Series Title: Proceedings of SPIE Web of Science ID: WOS:001254158700015.

\bibitem{halkonEstablishingCorrectionSolutions2021}
B.~J. Halkon and S.~J. Rothberg, ``Establishing correction solutions for scanning laser {Doppler} vibrometer measurements affected by sensor head vibration,'' {\em Mechanical Systems and Signal Processing}, vol.~150, p.~107255, Mar. 2021.
\newblock Num Pages: 21 Web of Science ID: WOS:000591667300003.

\bibitem{venkatakrishnanTwoaxisscanningLaserDoppler2002}
K.~Venkatakrishnan, B.~Tan, and B.~K.~A. Ngoi, ``Two-axis-scanning laser {Doppler} vibrometer for precision engineering,'' {\em Optics and Lasers in Engineering}, vol.~38, pp.~153--171, Oct. 2002.
\newblock Num Pages: 19 Web of Science ID: WOS:000176159400005.

\bibitem{zhuNumericalJitterEstimation2020}
L.~Zhu, A.~Miyazawa, P.~Mukherjee, I.~Abd El-Sadek, K.~Oikawa, D.~Oida, and Y.~Yasuno, ``Numerical jitter estimation for swept source optical coherence tomography,'' in {\em Biomedical {Imaging} and {Sensing} {Conference} 2020} (T.~Yatagai, Y.~Aizu, O.~Matoba, Y.~Awatsuji, and Y.~Luo, eds.), vol.~11521, (Bellingham), p.~115210T, Spie-Int Soc Optical Engineering, 2020.
\newblock Num Pages: 3 Series Title: Proceedings of SPIE Web of Science ID: WOS:000589972200026.

\bibitem{shapiro2012physics}
J.~H. Shapiro and R.~W. Boyd, ``The physics of ghost imaging,'' {\em Quantum Information Processing}, vol.~11, no.~4, pp.~949--993, 2012.

\bibitem{qiu2023remote}
X.~Qiu, H.~Guo, and L.~Chen, ``Remote transport of high-dimensional orbital angular momentum states and ghost images via spatial-mode-engineered frequency conversion,'' {\em Nature Communications}, vol.~14, no.~1, p.~8244, 2023.

\bibitem{zeng2025tailoring}
Y.~Zeng, Y.~Li, W.~Zhang, Y.~Cai, and L.~Chen, ``Tailoring entanglement with a symmetry: {Anomalous} ghost diffraction,'' {\em Physical Review Applied}, vol.~24, no.~5, p.~054002, 2025.

\bibitem{erkmen2012computational}
B.~I. Erkmen, ``Computational ghost imaging for remote sensing,'' {\em Journal of the Optical Society of America A}, vol.~29, no.~5, pp.~782--789, 2012.

\bibitem{yuanPlugandPlayAlgorithmsLargeScale2020}
X.~Yuan, Y.~Liu, J.~Suo, and Q.~Dai, ``Plug-and-{Play} {Algorithms} for {Large}-{Scale} {Snapshot} {Compressive} {Imaging},'' in {\em 2020 {IEEE}/{CVF} {Conference} on {Computer} {Vision} and {Pattern} {Recognition} ({CVPR})}, pp.~1444--1454, June 2020.
\newblock ISSN: 2575-7075.

\bibitem{ip2008coherent}
E.~Ip and J.~M. Kahn, ``Coherent detection in optical fiber systems,'' {\em Optics Express}, vol.~16, no.~2, pp.~753--791, 2008.

\bibitem{taylor2009phase}
M.~G. Taylor, ``Phase estimation methods for optical coherent detection using digital signal processing,'' {\em Journal of lightwave technology}, vol.~27, no.~7, pp.~901--914, 2009.

\bibitem{kikuchi2015fundamentals}
K.~Kikuchi, ``Fundamentals of coherent optical fiber communications,'' {\em Journal of lightwave technology}, vol.~34, no.~1, pp.~157--179, 2015.

\bibitem{fink1975coherent}
D.~Fink, ``Coherent detection signal-to-noise,'' {\em Applied Optics}, vol.~14, no.~3, pp.~689--690, 1975.

\bibitem{liuMicrovibrationModesReconstruction2022}
S.~Liu, C.~Deng, C.~Wang, Z.~Bo, S.~Han, and Z.~Lin, ``Microvibration {Modes} {Reconstruction} {Based} on {Micro}-{Doppler} {Coincidence} {Imaging},'' {\em Ieee Transactions on Geoscience and Remote Sensing}, vol.~60, p.~2008316, 2022.
\newblock Num Pages: 16 Place: Piscataway Web of Science ID: WOS:000896692300011.

\bibitem{qiMultimodeCoherentDetection2026}
J.~Qi, S.~Liu, C.~Deng, C.~Wang, Z.~Bo, Y.~Gui, and S.~Han, ``Multi-mode {Coherent} {Detection} {Ghost} {Imaging} {Lidar} and {Vibration}-{Mode} {Imaging},'' Jan. 2026.
\newblock arXiv:2601.13703 [physics].

\bibitem{kennedyParticleSwarmOptimization1995}
J.~Kennedy and R.~Eberhart, ``Particle swarm optimization,'' in {\em 1995 {Ieee} {International} {Conference} on {Neural} {Networks} {Proceedings}, {Vols} 1-6}, (New York), pp.~1942--1948, IEEE, 1995.
\newblock Num Pages: 7 Web of Science ID: WOS:A1995BF46H00374.

\bibitem{bobinMorphologicalComponentAnalysis2007}
J.~Bobin, J.-L. Starck, J.~M. Fadili, Y.~Moudden, and D.~L. Donoho, ``Morphological {Component} {Analysis}: {An} {Adaptive} {Thresholding} {Strategy},'' {\em IEEE Transactions on Image Processing}, vol.~16, pp.~2675--2681, Nov. 2007.

\end{thebibliography}

\newpage

\section{Biography Section}

\begin{IEEEbiography}[{\includegraphics[width=1in,height=1.25in,clip,keepaspectratio]{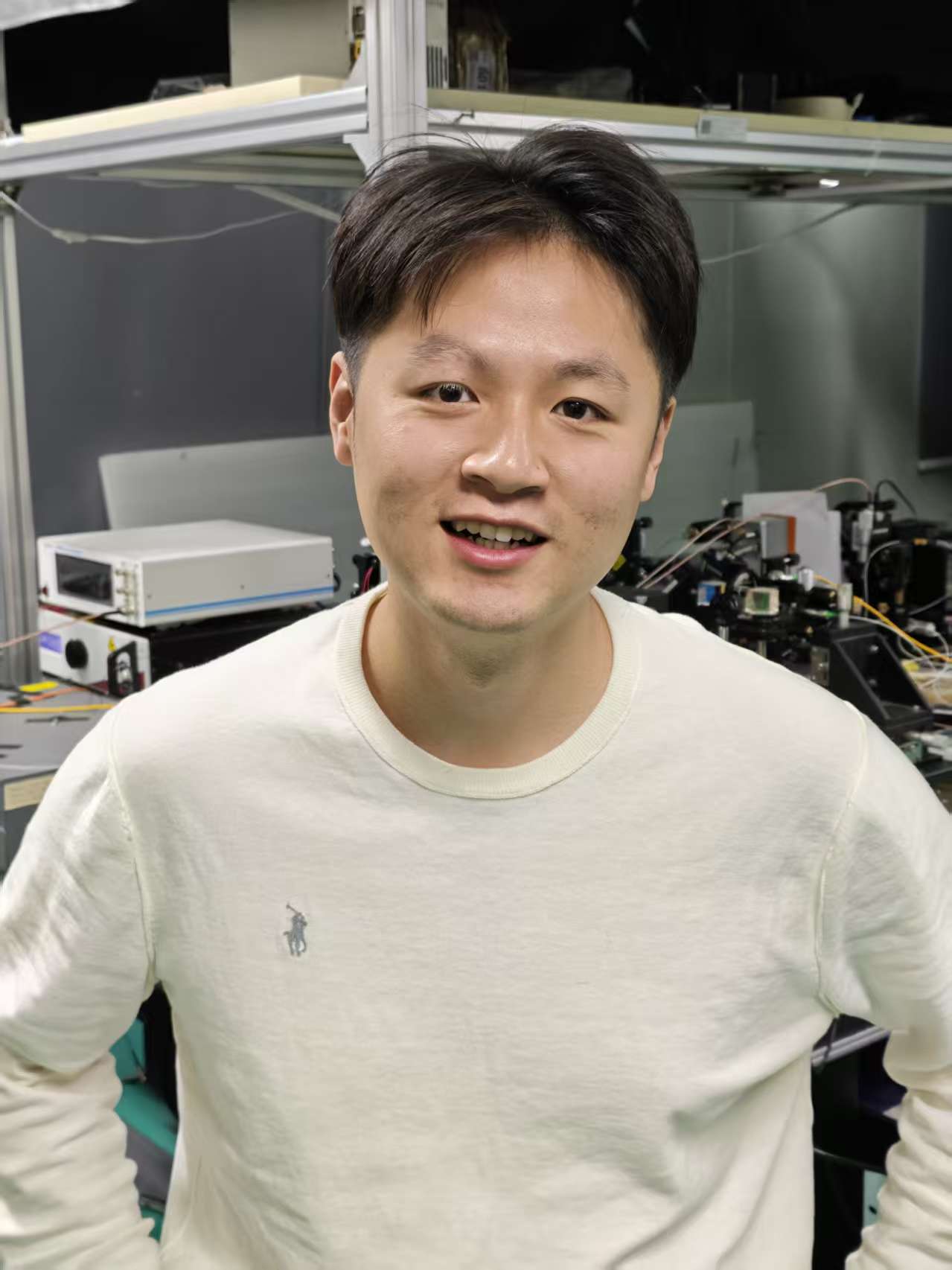}}]{Shuang Liu}
 received the Ph.D degree in Electrical Engineering from the University of Sydney in 2023. He is currently a postdoc in Optical Engineering with the Wangzhijiang Innovation Center for Laser, Aerospace Laser Technology and System Department, Shanghai Institute of Optics and Fine Mechanics, Chinese Academy of Sciences. His research interests include GI lidar signal processing and modeling of vibration targets.
\end{IEEEbiography}

\begin{IEEEbiography}[{\includegraphics[width=1in,height=1.25in, clip,keepaspectratio]{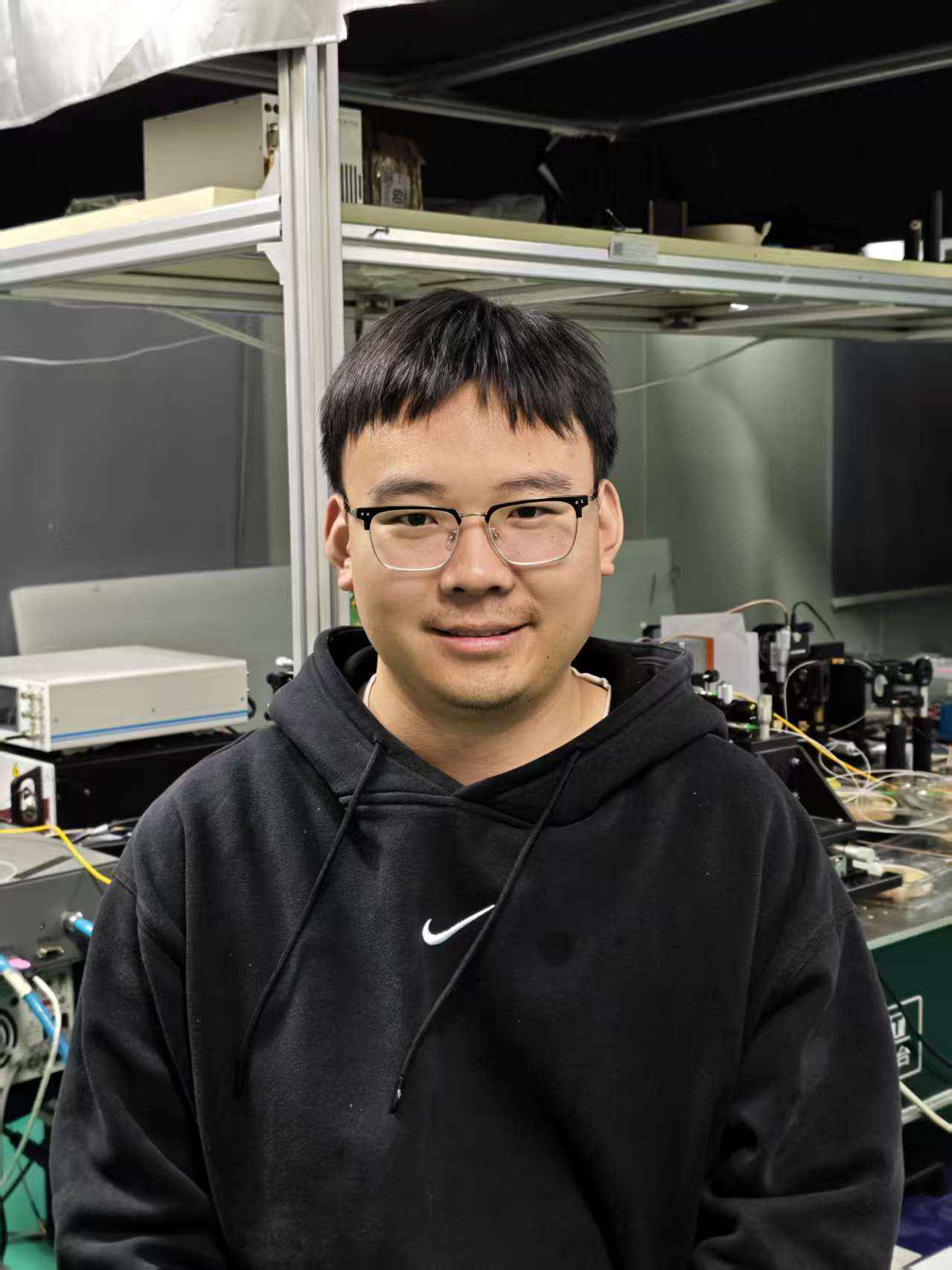}}]{Jinquan Qi} received the B.S. degree in Optoelectrical Information Science and Engineering from East China Jiaotong University in 2019. He is currently pursuing the Ph.D. degree in Optical Engineering with the Wangzhijiang Innovation Center for Laser, Aerospace Laser Technology and System Department, Shanghai Institute of Optics and Fine Mechanics, Chinese Academy of Sciences. His research interests include ghost imaging, coherent detection, and vibration mode imaging.
\end{IEEEbiography}

\begin{IEEEbiography}[{\includegraphics[width=1in,height=1.25in, clip,keepaspectratio]{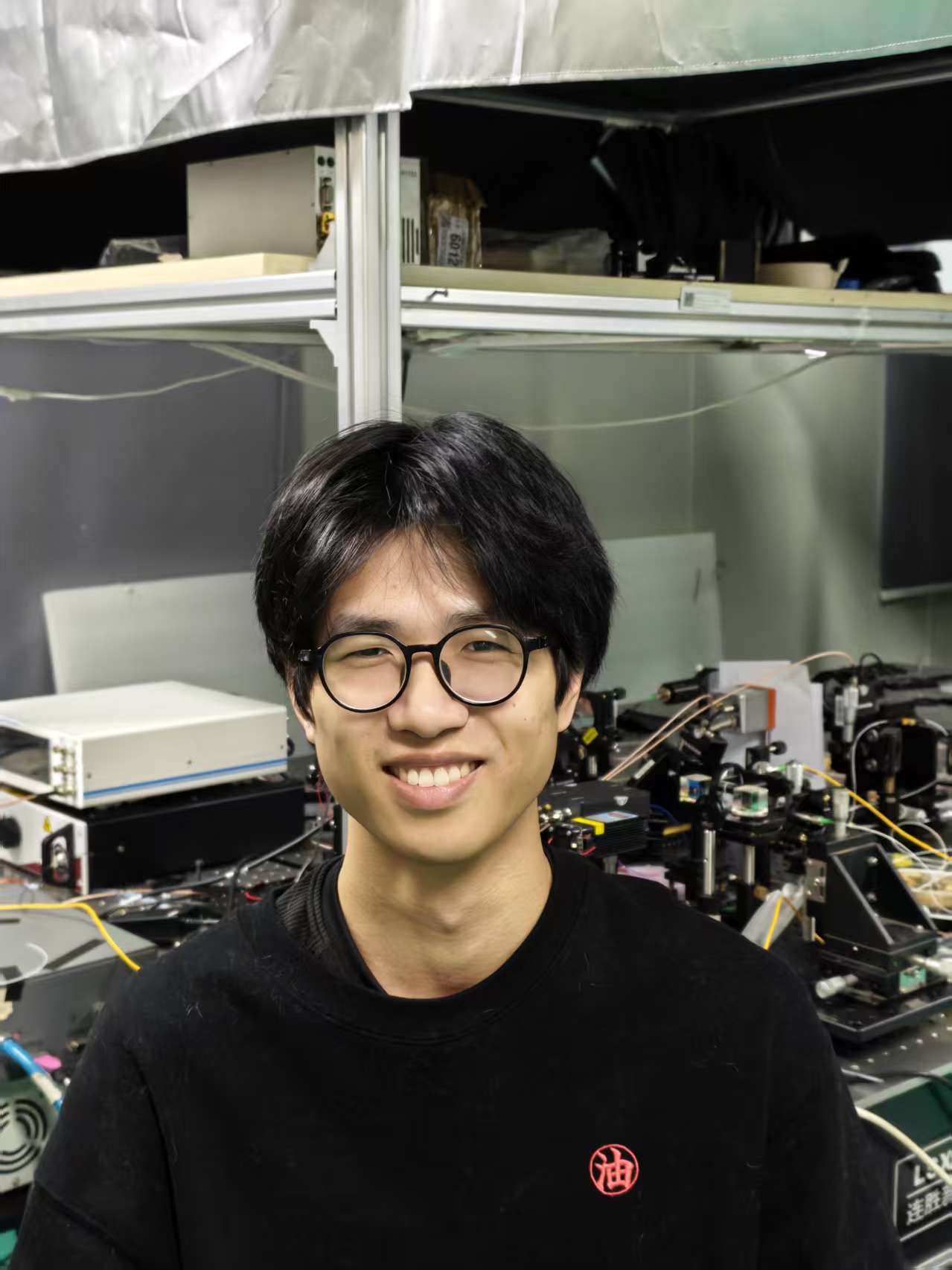}}]{Chaoran Wang} received the B.S. degree in Applied Physics from the School of Physics, Dalian University of Technology in 2021. He is currently pursuing the Ph.D. degree in Optical Engineering with the Wangzhijiang Innovation Center for Laser, Aerospace Laser Technology and System Department, Shanghai Institute of Optics and Fine Mechanics, Chinese Academy of Sciences.
His research interests include ghost imaging, stochastic partial differential equations, and random matrix theory.
\end{IEEEbiography}

\begin{IEEEbiography}[{\includegraphics[width=1in,height=1.25in, clip,keepaspectratio]{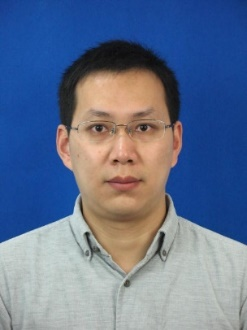}}]{Chenjin Deng} received his Ph.D. degree in Optical Engineering from the University of Chinese Academy of Sciences (UCAS) in 2018. He is currently an Associate Researcher with the Wangzhijiang Innovation Center for Laser, part of the Aerospace Laser Technology and System Department at the Shanghai Institute of Optics and Fine Mechanics (SIOM), Chinese Academy of Sciences (CAS). His research interests include imaging LiDAR systems and signal processing.
His research interests include ghost imaging, stochastic partial differential equations, and random matrix theory.
\end{IEEEbiography}

\begin{IEEEbiography}[{\includegraphics[width=1in,height=1.25in, clip,keepaspectratio]{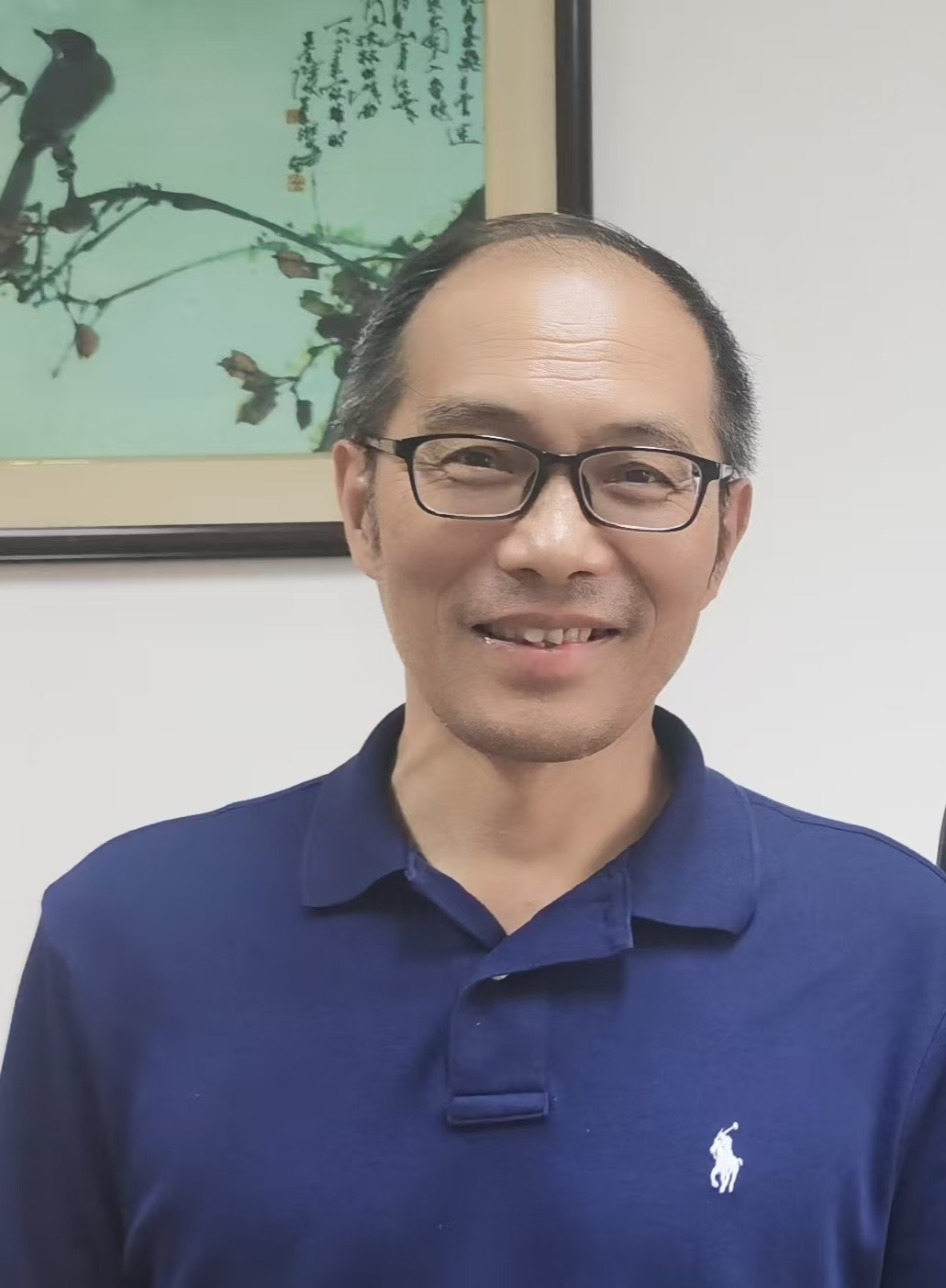}}]{Shensheng Han} received the Ph.D. degree in physics from University of Science and Technology of China in 1990, and completed his postdoctoral research at the Shanghai Institute of Optics and Fine Mechanics, Chinese Academy of Sciences (CAS), in 1992.
His research interests include strongly coupled plasma physics, quantum imaging and its applications, and novel X-ray imaging technologies.
\end{IEEEbiography}
\vspace{11pt}

\vfill

\end{document}